\documentclass[12pt]{article}
\usepackage{amsmath}
\usepackage{amssymb}
\usepackage{amsfonts}
\usepackage{amsthm}
\usepackage{paralist}
\usepackage{mathtools}
\usepackage{graphicx}
\usepackage{rotating}
\usepackage{setspace}
\usepackage{lscape}
\usepackage{array}
\usepackage{booktabs}
\usepackage{multirow}
\usepackage{caption}
\usepackage{footmisc}
\usepackage[authoryear,longnamesfirst,sort]{natbib}
\usepackage{bm}
\usepackage{verbatim}
\usepackage[margin=1in]{geometry}

\usepackage[hidelinks]{hyperref}

\usepackage{tabularx,booktabs} 
\usepackage[table]{xcolor}

\newcolumntype{L}[1]{>{\hsize=#1\hsize\raggedright\arraybackslash}X}%
\newcolumntype{R}[1]{>{\hsize=#1\hsize\raggedleft\arraybackslash}X}%

\theoremstyle{definition}

\DeclareMathOperator*{\Cov}{Cov}

\title{A Machine Learning Approach to Improving Occupational Income Scores\footnote{We are grateful for feedback from Ran Abramitzky, Brian Beach, Hoyt Bleakley, Melissa Dell, James Feigenbaum, Ekaterina Jardim, Bruce Sacerdote, James Siodla, Randall Walsh, and Eugene White, as well as seminar participants at the University of Michigan, Harvard University, the College of William \& Mary, Marquette University, and Clemson University. We are also grateful for comments from participants at the following conferences: WEAI, the Northeast Ohio Economics Workshop, SEA, the Pacific Northwest Labor Day conference at UW Seattle, and the North American Summer Meeting of the Econometric Society. Grant Goehring provided excellent research assistance. All errors are our own.}}
\date{ \today }
\author{ Martin Saavedra \\ Department of Economics \\ Oberlin College \\ \\ Tate Twinam \\ Department of Economics \\ College of William \& Mary}

\begin{document}

\maketitle

\begin{abstract}
Historical studies of labor markets frequently lack data on individual income. The occupational income score (OCCSCORE) is often used as an alternative measure of labor market outcomes. We consider the consequences of using OCCSCORE when researchers are interested in earnings regressions. We estimate race and gender earnings gaps in modern decennial Censuses as well as the 1915 Iowa State Census. Using OCCSCORE biases results towards zero and can result in estimated gaps of the wrong sign. We use a machine learning approach to construct a new adjusted score based on industry, occupation, and demographics. The new income score provides estimates closer to earnings regressions. Lastly, we consider the consequences for estimates of intergenerational mobility elasticities.
\end{abstract}

\noindent
{\bf JEL codes}: C21, J71, N32 \\
{\bf Keywords}: OCCSCORE, occupational income score, LIDO score, machine learning, lasso, non-classical measurement error, occupation, earnings gaps \\

\thispagestyle{empty}
\onehalfspacing

\pagebreak

\section{Introduction}

Before 1940, data on individual wages and education are not available in the U.S.\ Census. Consequently, occupation is often the only measure of labor market outcomes available to economic historians. Occupation is a categorical variable; however, many economists use occupational indices as continuous measures of historical labor market outcomes. One popular example is the 1950 occupational income score (OCCSCORE), which is roughly the median income of an occupation in 1950.\footnote{The 1950 OCCSCORE is the weighted mean of median male and median female earnings by occupation.} Originally developed by \citet{sobek1995comparability} as a continuous measure of occupational earnings potential, Sobek acknowledged that ``Although the income score is derived from individual-level data, it should not be interpreted as actual income.'' Since then, occupational income scores have been used to examine labor market outcomes going as far back as 1850, and studies using this approach have been published in numerous top journals in economics and other fields.\footnote{See section \ref{litreview} for examples.} 

Although occupational income scores are a reasonable proxy for occupational status, it is unclear how much bias this measurement error induces if researchers are primarily interested in earnings. Additionally, it is unclear if 1950 occupational income scores are good measures of income when examining Censuses several decades before 1950. While this potential bias has been acknowledged in the literature, only a handful of attempts have been made to quantify and diagnose the impact on inference.\footnote{The most direct example is \cite{inwood2019occupational}, which addresses the performance of occupational income scores in the context of immigrant assimilation in Canada. We discuss studies that have critically examined OCCSCOREs in section \ref{litreview}.} In this study, we attempt to measure this bias and examine how much it can be mitigated through adjustments to occupational income scores based on demographic and geographic variables available in all U.S.\ Censuses dating back to 1850.

We first discuss the potential biases introduced by the use of occupational income scores, decomposing these into two types. The first type of bias is due to changes in the earnings accruing to different occupations over time; these are difficult to measure due to a lack of historical income data. The second type of bias arises from within-occupation earnings differences across individuals, possibly due to gender, race, age, industry, location, or other factors.\footnote{While we dichotomize these biases, they can naturally interact, as we discuss further in section \ref{lido}.} We consider how these issues may affect the performance of the standard OCCSCORE when used as a proxy for income in an earnings regression, and discuss ways in which we may be able to improve upon OCCSCORE for this purpose.

Next, we analyze the performance of occupational income scores empirically. Because it is difficult to make historical data better, we proceed by making modern data worse.\footnote{Our approach is in the spirit of \cite{romer1986spurious}, who shows that excess volatility in unemployment time series during the pre-war era is an artifact of the interpolation methods used before the Current Population Survey. Applying the same interpolation methods to unemployment data during the post-war period results in similar levels of volatility.} We generate 2000-based occupational income scores and examine how well they predict income in the decades between 1950 and 2000. We then use this index to examine race and gender disparities from 1950 through 2000 and compare these to the true gaps estimated using actual earnings data. Finally, we use a technique from machine learning, cross-validated lasso regressions, to construct a new income score based on occupation, industry, race, sex, age, and state of residence. We compare estimated earnings gaps based on our lasso-adjusted industry, demographic, and occupation (LIDO) scores to those generated using OCCSCORE and true earnings. 

We find that although OCCSCORE is correlated with income even for Censuses five decades removed from the base year, earnings gaps are attenuated when using OCCSCORE as a proxy for income. We find that adjusting OCCSCORE by race, sex, age, industry, and geography reduces this bias. Estimated race and gender earnings gaps in 1915 Iowa using true earnings data are sizable and negative; however, when using standard OCCSCORE as a proxy for earnings, the racial earnings gap is underestimated by almost half and the gender earnings gap is statistically significant and of the wrong sign. Our LIDO score yields race and gender earnings gaps close to the true values. Finally, we conduct an analysis of OCCSCORE-induced bias in measures of intergenerational income transmission. This analysis is based on father-son pairs linked from the 1880 decennial Census to the 1850, 1860, 1900, 1910, 1920, and 1930 decennial Censuses. In this setting, we find that standard OCCSCOREs and our alternative scores provide similar estimates for white males. However, transmission coefficients for black males are larger for the LIDO score than the OCCSCORE. We conclude with recommendations for future research in economic history.

\section{Previous Literature}\label{litreview}

The occupational income score (OCCSCORE) was developed by Matthew Sobek and IPUMS for the purpose of representing ``the material rewards accruing to persons in different occupations'' \citep{sobek1995comparability}. It provides a continuous alternative to coarse occupational groupings and generates results that are more comparable to earnings regressions common in modern labor economics. To understand how researchers use this variable, we searched for papers containing either ``OCCSCORE'' or ``Occupational Income Score'' in top general interest journals and top field journals in labor economics and economic history. This search yielded the 25 papers listed in the Online Appendix.\footnote{This search included articles in the following journals: \emph{American Economic Journal: Applied Economics}; \emph{American Economic Journal: Economic Policy}; \emph{American Economic Review}; \emph{Explorations in Economic History}; \emph{Journal of Economic History}; \emph{Journal of Human Resources}; \emph{Journal of Labor Economics}; \emph{Quarterly Journal of Economics}; \emph{Review of Economics and Statistics}, and \emph{The Review of Economic Studies}.} 

Most of the articles have been published within the last decade, with a median publication year of 2014. Sixteen use the log of occupational income score as a dependent variable, and consequently, we focus our empirical analysis on the log of occupational income score. Of these 25 papers, only four adjust occupational income scores by any demographic variables.\footnote{The occupational earnings measure used by \cite{collins2014selection} varies by race and region; \cite{angrist2002sex} varies by age and sex; and \cite{collins2000african} presents results using both an unadjusted and a race-adjusted OCCSCORE.} Typically, these papers analyze historical Census data for which income or wage data are not available. Some of the papers, however, present occupational income score along with wage/income data as an alternative measure of socioeconomic status (see \cite{stephens2014compulsory} or \cite{chin2005long}). Some papers attempt to reduce bias by limiting the sample to a particular demographic group, typically white males (see \cite{bleakley2010malaria}). These papers often examine intergenerational mobility, racial and ethnic SES gaps, migrant selection, and the effects of schooling or health interventions.

This list underestimates how many researchers use occupational income scores or similar measures. Other papers in these journals may have used average or median income/wages by occupation as a dependent variable but do not refer to the variable as an occupational income score. \cite{inwood2019occupational} provide additional examples of older studies relying on occupational income scores. These scores are also used in other fields, especially sociology. A Google Scholar search for research articles containing ``OCCSCORE'' or ``occupational income score'' currently yields 291 articles. 

A number of authors have expressed concern about possible bias due to the use of OCCSCORE. \cite{abramitzky2014nation} point out that using OCCSCORE allows one to measure native-immigrant earnings convergence between occupations but not within occupation. Using 1970 and 1980 Census data, they estimate that using OCCSCORE captures at least 30\% of total earnings convergence between these two groups. Some authors, recognizing the issues we address here, have constructed occupational income scores adjusted for relevant covariates. For example, \cite{bailey2006wage} construct average occupational wages across sex-race-industry-region cells in 1940 to analyze the wage gains of black women between 1910 and 1940. \cite{inwood2019occupational} question the accuracy of results based on occupational income scores when examining immigrant assimilation. Utilizing Canadian Census data from 1911-1931, they construct income scores specific to different immigrant groups and show that these may outperform the standard OCCSCORE when attempting to approximate the results from a true earnings regression.

\section{Constructing an Adjusted Occupational Income Score}\label{lido}

OCCSCORE works by capturing variation in individual earnings due to occupational sorting. There are two key limitations to this approach. First, it fails to capture variation in earnings within occupation; in particular, it ignores any variation due to individual characteristics such as race, gender, and age, or individual circumstances, such as industry of employment and geographic location. Second, since it is based on 1950 data, it fails to reflect changes in occupational earnings rankings or changes in how individual characteristics affect within-occupation earnings over time. Without detailed pre-1950 earnings data, it is not possible to address this latter problem. However, using 1950 data, we can address the former problem, and this is our primary goal.

A simplified example will help to illustrate these issues. Suppose our goal is to estimate the gender earnings gap in 1950 using an occupational income score based on data from 2000. We divide the population into two ``occupations''--blue-collar and white-collar. Among both men and women, white-collar workers (on average) earn more than blue-collar workers. Among both blue-collar and white-collar workers, men (on average) earn more than women. The ratio of average female income to average male income is 0.57 in our 2000-based sample; this is the true earnings gap in 2000. Suppose we now compute average earnings by occupation, ignoring gender; we find that average earnings for white-collar and blue-collar workers are \$38,313 and \$22,762, respectively. These values will form the 2000-based OCCSCORE for these occupations. We can then apply these values to the distribution of male and female workers across these occupations in 1950 to derive an estimate of the gender earnings gap. We find a female/male earnings ratio of 1.06, indicating that female workers, on average, earned \textit{more} than male workers in 1950. The true earnings ratio in 1950 is 0.52. Furthermore, if we instead apply our 2000-based OCCSCORE to the year 2000 sample, we find an earnings ratio 1.12--far from the true value of 0.57, despite the fact that our 2000-based OCCSCORE is being applied to the same sample used to construct it.

What went wrong? The spurious sign-reversal on the gender earnings gap we observe in this example is a manifestation of Simpson's Paradox. While women earn less than men both overall and within both occupations in 2000, women are disproportionately likely to be white-collar workers rather than blue-collar workers.\footnote{In our 2000 sample, there are 2.4 white-collar female workers for each blue-collar female; the number is 0.9 for males. In 1950, women are evenly split across occupations, while there are only 0.6 white-collar male workers for each blue-collar male.} Thus, an income score based only on occupation greatly overstates womens' earnings relative to mens', both in 2000 and 1950, and the ignored within-occupation earnings differences are large enough to flip the sign of the gender earnings gap. A partial solution to this problem involves exploiting our knowledge of these within-occupation earnings differences in 2000. Instead of averaging earnings across all individuals in each occupation, we can average earnings within each gender-occupation cell, yielding a 2000-based OCCSCORE adjusted for gender differences. Applying this adjusted income score to the 1950 data yields a female-male earnings ratio of 0.54, much closer to the true value of 0.52. Why does the adjusted score not match the true value exactly? Both occupational earnings and within-occupation gender earnings differences are not static over time; our adjustment captures the correct values in 2000, ignoring the fact that these have evolved since 1950. Nonetheless, the adjustment is sufficient to reduce the bias of our 1950 earnings gap estimate from 104\% to only 4\%, despite the substantial changes in the labor market for women that occurred between 1950 and 2000. The remaining bias can only be corrected with actual historical data on the evolution of earnings; neither the standard OCCSCORE nor the alternative scores we develop below address this aspect of the problem.

The above example suggests that a score derived from average incomes conditioned on both occupation and a suite of common explanatory variables of interest should provide estimates closer to those of a true earnings regression. The most important commonly-available variables influencing labor market outcomes are industry, sex, race/ethnicity, age, and geographic location.\footnote{Another common demographic variable that could be used is an indicator for foreign-born status. However, we caution that this may be misleading. The composition of the foreign-born population in the U.S.\ in 1950 differs dramatically from that in earlier years such as 1900 and 1850 in both racial/ethnic makeup and human capital. Thus, adjusting on this variable in a given base year may lead to inaccurate results when applied to other years. \cite{inwood2019occupational} provide a more detailed discussion of this issue in the context of early $20^{\text{th}}$ century Canadian immigration patterns.\label{immigrationnote}} Any adjusted OCCSCORE should account for differences along these lines. While other variables may no doubt be important, we focus on these because they are consistently available across decennial Censuses, meaning that our adjusted score will be widely applicable. However, researchers examining particular questions or using non-Census data sources can use the method we propose below to construct scores adjusted by any relevant variables of interest.

The simplest and most general approach would be to adjust OCCSCORE in a fully nonparametric manner. For example, one could take an individual's OCCSCORE to be the median or mean income in a given base year for that individual's occupation within cells defined by their sex, age, race, state, and industry.\footnote{For example, \cite{angrist2002sex} constructs age- and sex-specific OCCSCOREs based on median income within cells. \cite{collins2014selection} compute income scores by occupation and region specifically for black men.} The advantage of this measure is that it allows for arbitrary interactions between all of the adjustment variables. The disadvantage of this approach is that stratifying on so many variables may result in small or empty cells, leading to excessively variable or missing OCCSCOREs for many individuals.\footnote{This limitation is partly due to the fact that we are currently constrained to the 1\% sample of the 1950 Census available through IPUMS. However, even with the full count data that will be available in the future, the issue of small/empty cells likely remains a problem. For example, about 15\% of working-age adults in 1900 are in an occupation, industry, age, state, race, and sex cell that is empty in the 1940 full count.}

An alternative involves a less flexible parametric approach. For example, one could regress income in a given base year on a series of occupation, sex, age, race, and geographic state indicator variables. The fitted coefficients could then be used to generate an adjusted OCCSCORE for each possible individual. This strategy is computationally simple and generates an adjusted OCCSCORE for all individuals. However, it likely misses many important interactions. For example, there is little reason to believe that early $20^{\text{th}}$ century earnings gaps between whites and blacks did not differ by region, nor does it seem likely that the age-earnings profile was the same across all occupations and industries.

Our approach aims to balance the need for a rich model of income determinants with the limitations imposed by the small number of observations available for some occupations. For a given base year, we compute lasso-adjusted industry, demographic, and occupation (LIDO) scores as follows. For each Census-classified industry, we regress log income on a set of demographic covariates for all individuals between the ages of 20 and 70 employed with positive earnings in that industry.\footnote{Industries follow the 1950 Census Bureau industrial classification system. Stratifying by industry eases the computation burden of the algorithm.} We use the lasso algorithm, which solves the standard least squares problem subject to a constraint on the sum of the absolute values of the model coefficients \citep{tibshirani1996regression}. This regularization approach controls the complexity of the model based on the importance of the predictors and the size and composition of the sample.

We allow for the following regressors: indicators for all occupations within the given industry, a polynomial for age, indicators for sex, race, and state of residence, and interactions between \begin{inparaenum}[(1)] \item sex and race, \item sex and region, \item occupation and sex, \item occupation and an indicator for white, \item Census region and an indicator for white, and \item Census region and an indicator for black. \end{inparaenum} In 1950, this results in a maximum of 654 possible covariates for the industry with the largest number of represented occupations (educational services). In general, the number of possible covariates is large relative to the sample size for each industry, and in some cases it may exceed the number of observations. The lasso algorithm shrinks coefficients depending on their relative importance, with the constraint forcing the coefficients on the least relevant predictors to zero. The sparsity induced by the lasso depends on the choice of tuning parameter $\lambda$ for each particular industry (described further below).

Like OCCSCORE, the LIDO score still relies on variation in earnings across occupations. However, it additionally allows for individual characteristics to influence within-occupation earnings. Income for a given occupation can vary depending on the particular industry in which an individual works; it can also vary in flexible ways with race, sex, and geographic region. The age profile of earnings can also differ by industry. This generates scores that more closely proxy for true income than the interaction-free regression approach described above. 

Our approach also avoids the small-cell overfitting problem that arises from the fully nonparametric approach; the lasso retains only the most relevant predictors of income differences, and the complexity of the model is scaled based on the availability of observations in each industry. The extent to which the lasso generates a sparse model depends on the choice of tuning parameter $\lambda$, which reflects the stringency of the constraint. Since the importance of different demographic factors likely varies by industry, a one-size-fits-all choice would be inappropriate. We instead use 10-fold cross-validation to select a $\lambda$ that minimizes out-of-sample mean squared error for each industry.

An important limitation of this approach is that all within-occupation earnings differences are based on 1950 data, and there is no guarantee that these earnings differences are stable over time. This limitation applies to the standard approach as well, since OCCSCORE also assumes that earnings gaps within occupation are stable over time, with the additional restriction that they are uniformly equal to zero. Without pre-1950 earnings data indexed to demographics, it is difficult to assess how this would affect results based on the LIDO score (or OCCSCORE) applied to earlier decades. Our analysis in section \ref{results} uses modern data to measure the extent to which changes in earnings gaps affect the reliability of our LIDO score relative to OCCSCORE.

\section{Results}\label{results}

In this section, we document the performance of the LIDO score relative to OCCSCORE using modern Census data.\footnote{Replication files are available on ICPSR \citep{occdata}.} Initially, we examine the stability of occupational earnings over time. We then estimate earnings regressions between 1950 and 2000 and measure the extent to which OCCSCORE causes bias when estimating race and gender earnings gaps. 

\subsection{Persistence of Occupational Income}\label{persist}
 
The occupational income score is a weighted average of the median earnings for males and females for each occupational category in 1950. This variable is likely a reasonable proxy for earnings in 1950, but it is unclear whether the relative earnings of occupations are sufficiently stable for it to remain an accurate proxy for income in earlier decades. If they are not, then even an adjusted version of OCCSCORE may perform poorly.

We test whether median earnings of an occupation accurately predict median earnings in the decades before the base year. We do this by constructing a 2000-based OCCSCORE and testing how well it predicts median earnings from past Censuses. If the 2000-based OCCSCORE successfully proxies for median earnings in 1950, then the 1950 OCCSCORE may be a reasonable proxy for median occupational earnings in 1900. 

\begin{figure}
\begin{center}
\caption{2000 Occupational Income Score and Median Income}
\label{fig:occscore2000a}
\scalebox{.25}{\includegraphics{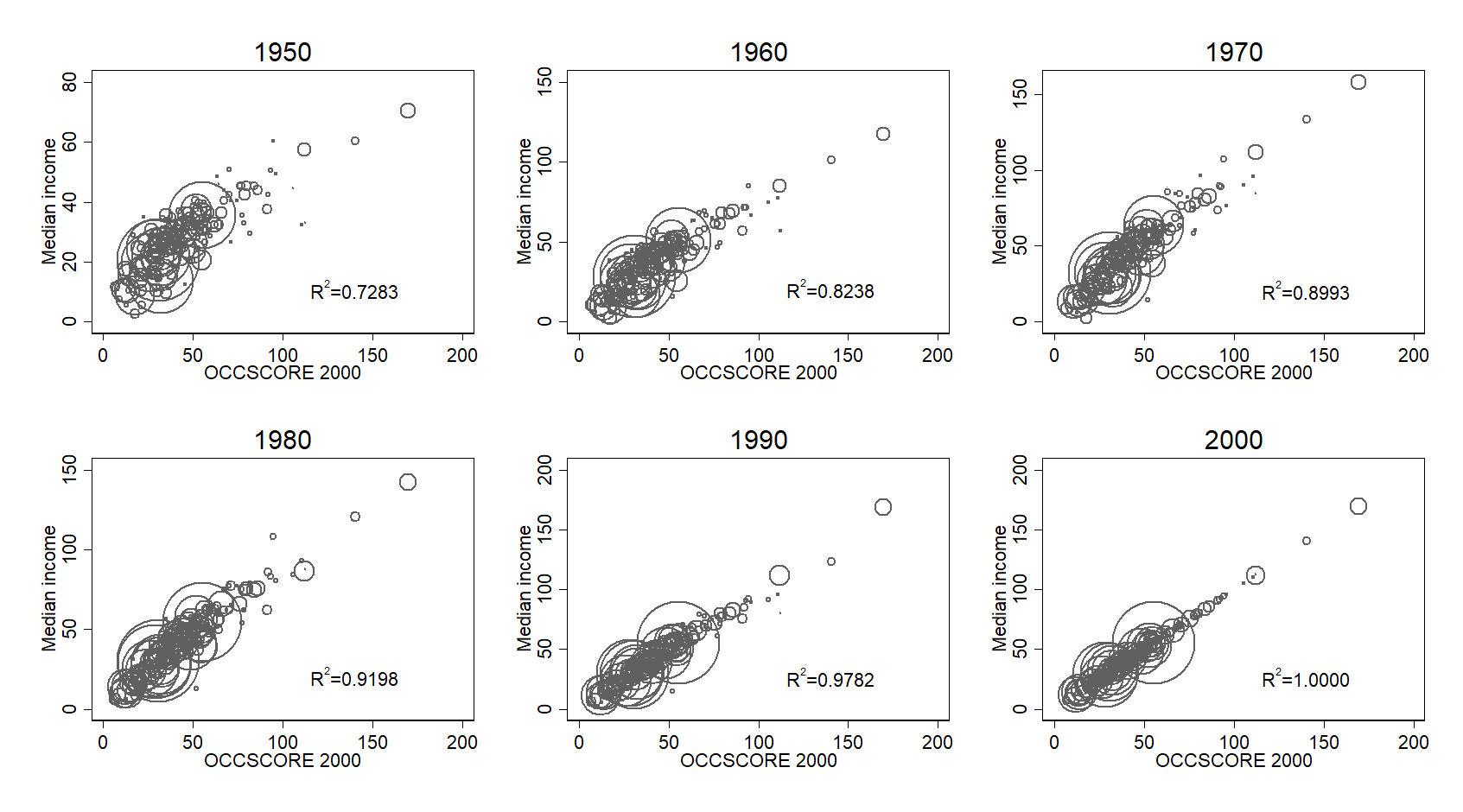}}
\end{center}
\footnotesize{ {\bf Notes:}  Median earnings for each occupation are from the 1\% sample of the U.S.\ Census. The 2000-based OCCSCOREs are the median earnings from the 2000 Census for individuals in each occupation. The size of each circle corresponds to the number of individuals in the occupational category during that Census year. Median earnings and OCCSCORE are measured in hundreds of 1950 dollars.}
\end{figure}

The results of this exercise can be seen in Figure \ref{fig:occscore2000a}. Each circle is an occupation weighted by the size of the occupational cell. The 2000 OCCSCORE perfectly predicts median earnings in 2000 by construction. For each decade removed from 2000, the $R^2$ decreases, implying that OCCSCORE is becoming worse as a proxy for median earnings. Even 50 years removed from the base year, $R^2=0.73$, implying that OCCSCORE remains a strong proxy for median earnings.  

To provide further evidence, we examine changes in the rank correlation of median occupational income between 1950 and 2000. Measured by Spearman's rank correlation coefficient, the correlation between occupational rankings in 1990 and 2000 is 0.97. While this declines over time, it does so gradually. Between 1950 and 2000, the correlation is 0.81, which is still very high. While we cannot examine how this correlation changes nationally in the decades before 1950, we can examine the correlation between occupational income in 1950 and that in 1915 Iowa using state Census data (described in section \ref{iowa}). Using this information, we find that the rank correlation between 1950 and 1915 occupational earnings in Iowa is 0.7. This analysis provides some validation for the use of occupational income scores, since it demonstrates that the earnings hierarchy of occupations is likely to be reasonably stable over time.

\subsection{Earnings Gap Bias}\label{bias_section}

While many economic historians use 1950-based occupational income score as a proxy for occupational status, some interpret it as a proxy for earnings, and then estimate models using data from pre-1950 Census years. We cannot directly test whether OCCSCORE produces coefficients similar to earnings regressions using pre-1950 national Census data. However, we can make modern data worse, so that the modern data suffers from the same problems as historical data. Here, we generate a 2000-based occupational income score and compare estimated racial and gender gaps from 1950-1990 with the true earnings gaps a researcher would have obtained by using actual earnings instead of the proxy. 

\begin{figure}
\begin{center}
\caption{Earnings Ratios Using Earnings, OCCSCORE, and LIDO Score}
\label{fig:gaps}
\scalebox{.3}{\includegraphics{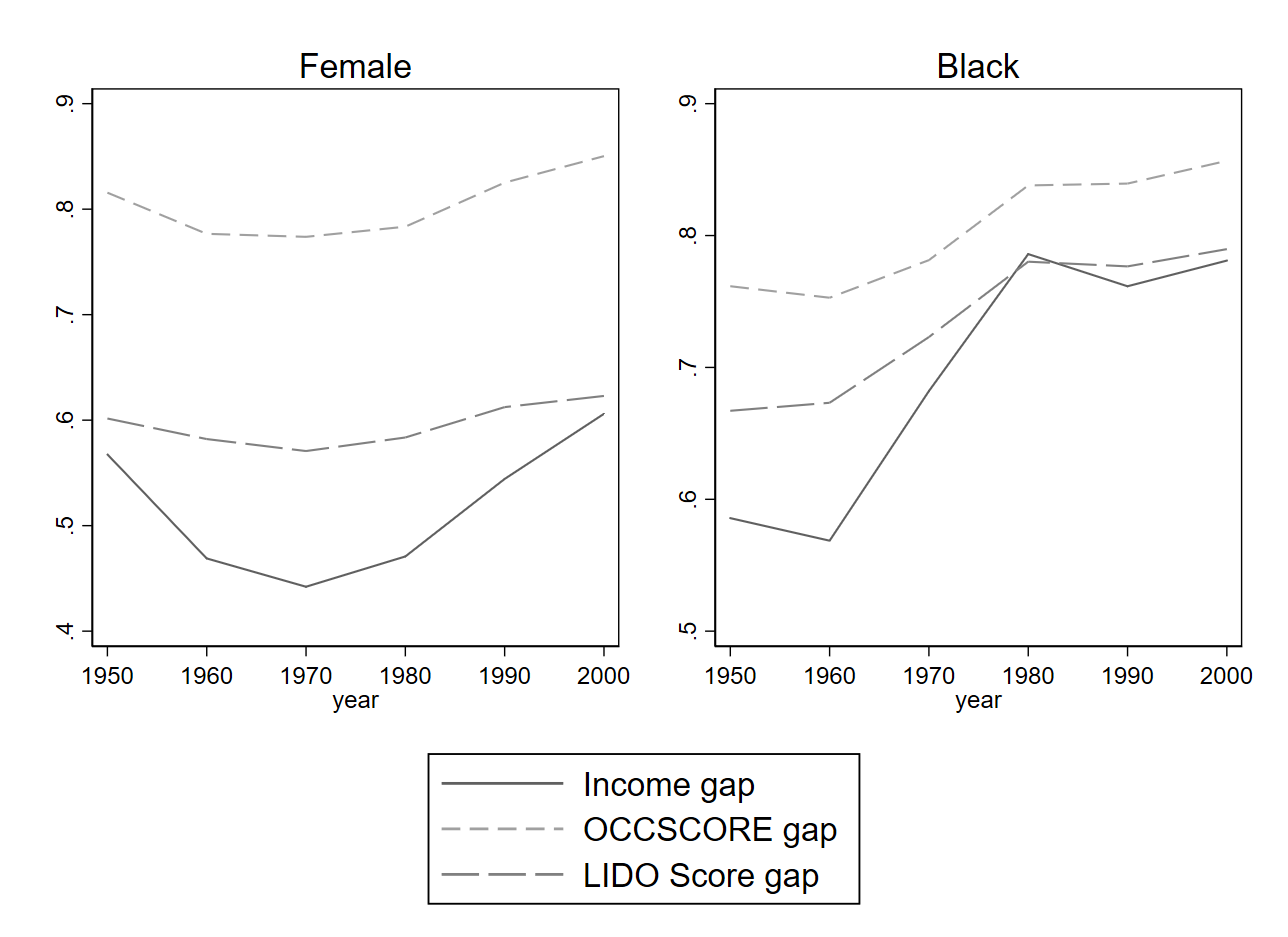}}
\end{center}
\footnotesize{ {\bf Notes:} The data are from IPUMS (see \cite{ruggles2015}). The graph displays the implied female/male and black/white income ratios from six earnings regressions. Note that the gaps are conditional on age, age squared, a dummy variable for U.S.-born, and state of residency. The female earnings gap is conditional on race, and the black/white earnings gap is conditional on sex. OCCSCORE uses a 2000-based occupational income score, whereas LIDO score is a 2000-based occupational income score constructed as described above. The sample is restricted to those between ages 25 and 65 who were in the labor force.}
\end{figure}

Figure \ref{fig:gaps} graphs the implied earnings gaps using three regressions. The first specification regresses the log of earnings on a set of dummies for state of residence, sex, race, and nativity. In addition to these dummy variables, the regression includes age and age squared. We run the regressions separately for every Census year from 1950 to 2000. Because we assume researchers would have used earnings instead of occupational income scores if earnings data were available, we treat these coefficients as the true parameters that researchers would like to estimate. The second regression uses the log of the 2000-based OCCSCORE instead of log earnings as the dependent variable. The dependent variable for the last regression is the log of the 2000-based LIDO score. For each regression, we restrict the sample to adults ages 25-65 who were in the labor force.

As expected, earnings gaps have declined for blacks since the 1950s and for women since the 1970s, and this is reflected in all three models. For all years, the coefficients on sex and race are of the same sign and statistically significant in all specifications, but the coefficients from the OCCSCORE specification suffer from attenuation bias. Using the LIDO score as the dependent variable reduces this bias, but does not eliminate it.  The earnings gap estimated using the LIDO score more closely mirrors the true earnings gap than the OCCSCORE estimates.\footnote{While the earnings gap estimated using the LIDO score is much closer to the true earnings gap in magnitude, the estimated trend in the earnings gap is very similar using both LIDO and OCCSCORE. This is unsurprising, since estimated changes in the male-female and black-white gap are driven by changes in occupational sorting regardless of which method is used.} 

Our estimates of the female/male earnings ratio are similar to the extant literature \cite[p.\ 62]{goldin1990gender}. The female/male earnings ratio declined between 1950-1960, after which the gender gap slowly narrowed. \cite{margo2016obama} provides Census estimates of the black/white earnings gap that are similar to ours. Black income increased relative to whites during the 1960s and 1970s, but the ratio has not narrowed significantly since the 1980s. \cite{smith1984race} estimates the black/white income gap by assigning each individual the average income of a race by sex by age group cell from the 1970 Census. These estimates, produced at least a decade before the IPUMS OCCSCORE variable was regularly in use, are in essence an adjusted OCCSCORE.

\section{Applications}\label{applications}

\subsection{Earnings Gaps in the 1915 Iowa Census}\label{iowa}

The analysis in Section \ref{results} shows that the LIDO score improves estimates of racial and gender earnings gaps in modern Census data. To assess whether this conclusion applies in a historical context, we exploit a rare source of pre-1950 income data, the 1915 Iowa State Census \citep{goldin2010iowa}.\footnote{This data was recently used to examine intergenerational mobility by \cite{feigenbaum2018multiple}.} This was the first Census in the U.S.\ to collect data on income prior to 1940. The sample contains records on 5.5\% of the urban population drawn from three of Iowa's largest cities: Des Moines, Dubuque, and Davenport. It also contains 1.8\% of the population of counties not containing a major city; the ten counties sampled span the geography of the state. This data allows us to compare racial and gender earnings gaps and the age-income profile estimated using OCCSCORE, LIDO score, and true earnings in a historical setting.

For the estimation, we restrict the sample to those between the ages of 20 and 70 and exclude those with missing occupation data or zero/missing earnings.\footnote{We also exclude those whose race is recorded as missing (19 observations) and those whose race is recorded as Mixed or Asian (5 observations).} The Census reports occupation categories according to the 1940 scheme. We crosswalked these with the 1950 scheme to match individuals in 1915 to their 1950 OCCSCORE.\footnote{In some cases, the 1940 scheme aggregated some occupations; for example, bookkeepers, accountants, and cashiers fall into one occupation category in 1940 but are disaggregated into three separate categories in 1950. There are 7 occupation categories in 1940 (out of 194 total) that cannot be matched uniquely to a 1950 occupation. We exclude individuals in these categories.} The final sample includes 15,201 individuals. We estimate the earnings gap between whites and blacks and men and women; approximately 1\% of the sample is black (196 obs) and 14\% of the sample is female (2,153 obs). We also estimate the age-earnings profile, urban-rural gap, and the native-foreign born gap.

In column (1) of the top panel of Table \ref{iowa_results}, we report the coefficients from a regression of log earnings on indicators for black, female, urban, and foreign-born, as well as a quadratic polynomial for age. Women and African-Americans earn less than white men and, as is typical, earnings increase with age but at a diminishing rate. In column (2), we replace log earnings with the log of the standard 1950 OCCSCORE. The black-white earnings gap coefficient declines by almost half. The gender earnings gap is positive. The age-earnings profile is attenuated, as is the gap between natives and immigrants, while the urban premium is overestimated.

Moving to column (3), we replace OCCSCORE with our 1950 LIDO score.\footnote{Because industry was not recorded in the Iowa State Census, we instead estimate our lasso-adjusted score by occupation (retaining all of the other predictors listed in section \ref{lido}).} Using this approach, the earnings gap for women is similar to that estimated using true earnings. The estimated earnings gap for blacks is slightly larger than the true value but closer in magnitude than the standard OCCSCORE estimate. The age-earnings profile, while still attenuated, is closer to the correct value, as is the urban premium, which was not explicitly incorporated in the construction of the LIDO score. The OCCSCORE and LIDO score estimates of the immigrant penalty are similar.\footnote{Since farm income is unusually heterogeneous, and farmers saw a substantial change in their occupational standing between 1850 and 1950, we examined our results excluding those with the 1950 occupational classification ``Farmers (owners and tenants).'' The LIDO score again yields estimates closer to the earnings regressions for the female gap and age-earnings profile, though it does not deliver a more accurate estimate of the black-white gap. Results are similar if we additionally exclude individuals whose occupational classification is ``Managers, officials, and proprietors (n.e.c.),'' a category that is problematic due to the aggregation of small business proprietors and chief executives of large corporations.}

\subsection{Estimates of Intergenerational Mobility}\label{mobility}

Labor economists often measure intergenerational mobility by regressing a son's socioeconomic status on his father's socioeconomic status:
\begin{equation}
I_i^{\mbox{son}}=\beta_0+\beta_1 I_i^{\mbox{father}}+u_i
\end{equation}
where $I_i^{\mbox{son}}$ is the log income of a son observed during adulthood, and $I_i^{\mbox{father}}$ is the log income of a father observed while the son was a child. The transmission coefficient $\beta_1$ is an elasticity typically between 0 and 1, with 1 representing perfect immobility between generations and 0 representing perfect mobility. Historical evidence on occupational mobility across generations relies heavily on occupational income scores instead of income for two reasons. First, to obtain data on fathers' and sons' labor market outcomes in the Census, one needs to link across Census years, which is typically only possible using given and surnames. Names do not become publicly available in the Census until 72 years after the Census year, meaning occupations are the only available labor market outcomes for both fathers and sons. Second, estimates of how intergenerational mobility have changed over time require data spanning at least three generations, implying that such estimates must make use of historical data.

As \cite{solon1989biases, solon1992intergenerational} has highlighted, measurement error in the dependent variable has the potential to bias intergenerational mobility estimates in favor of greater mobility. Let $e_i^{\mbox{son}}=I_{i}^{\mbox{son}}-\tilde{y}_i^{\mbox{son}}$ and $e_i^{\mbox{father}}=I_{i}^{\mbox{father}}-\tilde{y}_i^{\mbox{father}}$ be the measurement error from using an occupational index (either OCCSCORE or LIDO score) for the son and father, respectively. Then researchers estimate:
\begin{equation}
\label{eq:trans}
\tilde{y}_i^{\mbox{son}}=\beta_0+\beta_1\tilde{y}_i^{\mbox{father}}+\underbrace{\beta_1e_i^{\mbox{father}}-e_i^{\mbox{son}}+u_i}_{\epsilon_i}.
\end{equation}

This regression differs from the model in Section 3 since OCCSCORE appears on both the left-hand and right-hand side of the regression. The measurement errors $e_i$ are likely to be smaller if one uses a demographically-adjusted LIDO score instead of OCCSCORE, since racial, age, and regional differences in occupational earnings will not be captured in $e_i$. However, when using OCCSCORE, some of the measurement error is likely to cancel out since $e_i^{\mbox{son}}$ is positively correlated with $e_i^{\mbox{father}}$. Our estimate of the transmission coefficient will be biased if $\Cov(\tilde{y}_i^{\mbox{father}},\beta_1e_i^{\mbox{father}}-e_i^{\mbox{son}})\neq 0$. If there is little intergenerational mobility, in which case $\beta_1$ is close to 1, and if the son's measurement error is highly correlated with the father's measurement error, then the second term of covariance is close to zero. Alternatively, suppose $e_i^{\mbox{son}}=\tilde{\beta}_0+\tilde{\beta_1}e_i^{father}+\upsilon_i$, where $\upsilon_i$ is independent of all other variables. The transmission coefficient $\tilde{\beta_1}$ reflects that fathers who earn above average within their occupations are likely to have sons who earn above average within occupations. Then, $\Cov(\tilde{y}_i^{\mbox{father}},\beta_1e_i^{\mbox{father}}-e_i^{\mbox{son}})=\Cov(\tilde{y}_i^{\mbox{father}},\beta_1 e_i^{\mbox{father}}-\tilde{\beta_1} e_i^{\mbox{father}})$. Thus, the bias from estimating equation \ref{eq:trans} using OCCSCOREs will be small so long as the transmission in overall income from father to son is similar to the transmission of excess income within occupation. For these reasons, using occupational income scores instead of income should lead to a smaller amount bias in this context.

In Table \ref{table:inter}, we provide estimates of intergenerational mobility using the IPUMS linked data sets. These data link the 1\% samples of the 1850, 1860, and 1900-1930 Censuses to the 1880 complete count. The sample is restricted to those who during the first Census year were children of the household head, no older than 15 years old, and male. We regress the log of a son's OCCSCORE during the second Census year on the log of the father's OCCSCORE during the first Census year, and then repeat the regression using LIDO score. To make the estimates comparable, we restrict the sample to father-son pairs in which neither the father nor the son has a missing LIDO score. The resulting coefficients are elasticities with higher coefficients implying occupational immobility. Row 1 of columns (2) and (4) of Table \ref{table:inter} are replications of the estimates in row 7 of Table 3 of \citet{olivetti2015name}, but restricting the sample to those with non-missing LIDO scores. We present estimates for whites using all samples, and for blacks using samples in which both the father and son are observed in the postbellum era.

The intergenerational mobility estimates for whites are similar for both measures. However, the OCCSCORE estimates suggest that blacks had twice the intergenerational mobility of whites ($\beta_1$ closer to zero). The LIDO score estimates suggest that black intergenerational mobility was much closer to white intergenerational mobility than previously thought. In fact, between 1880-1900 and 1880-1910, blacks had less intergenerational mobility than whites. Since we do not observe true incomes in historical censuses, we do not know what the true intergenerational mobility was. However, that the LIDO score and OCCSCORE disagree for blacks suggests we should should treat any occupational based measures of income mobility with caution.

\section{Conclusion}

Using modern Census data, we find that median earnings within a given occupation are highly correlated over time. This implies that the occupational income score is a reasonable proxy for occupational status even when used with historical Census data. However, much of modern labor economics focuses on earnings regressions rather than occupational status, and we find that occupational income scores systematically underestimate income gaps due to race and gender. We construct a new lasso-adjusted industry, demographic, and occupation (LIDO) score which flexibly accounts for differences in earnings across race, gender, age, state, occupation, and industry (variables available in every Census going back to 1850). Our alternative score provides estimates closer to earnings regressions. We have made this alternative score available online.\footnote{It can be found at \url{http://www2.oberlin.edu/faculty/msaavedr/lido.html}.} 

To examine the performance of the LIDO score in a historical context, we exploit the 1915 Iowa State Census, which collected data on both occupation and earnings. We find that estimated race and gender earnings gaps in 1915 Iowa using true earnings are sizable; however, when using standard OCCSCORE as a proxy, the racial earnings gap is attenuated by almost half and the gender earnings gap is incorrectly signed. Our LIDO score yields earnings gaps close to their true values. We also use the LIDO score to measure intergenerational income transmission. This analysis is based on father-son pairs linked across the 1850-1930 decennial Censuses. In this setting, we find that standard OCCSCOREs and LIDO scores perform similarly for white males because measurement errors for fathers and sons are likely to be correlated. However, transmission coefficients are attenuated for black men, suggesting a lower rate of intergenerational mobility than previously thought.

Our results suggest that future research in economic history and other fields may benefit from the use of LIDO scores if researchers are interested in earnings regressions. When should researchers continue to use standard occupational income scores? As \cite{inwood2019occupational} point out, unadjusted OCCSCOREs may be a reasonable proxy for total earnings over the life cycle. A recent college graduate may have a low adjusted OCCSCORE, since young professionals are below their peak occupational earnings, whereas the standard OCCSCORE would assign workers of all ages within the occupation the same earnings.\footnote{\cite{corak2006poor} and \cite{feigenbaum2015intergenerational} observe that age-related variation in earnings could bias estimates of intergenerational mobility coefficients downward.} Race and gender earnings gaps may also shrink over the life-cycle. Without linked Census data, it is impossible to know whether OCCSCOREs or LIDO scores provide a better proxy for lifetime wealth. Although linked Census data does exist, it does not cover the 1950-2000 periods in which earnings data are available.\footnote{The only linked Census data we are aware of that includes pre-1950 information on total income is based on the Iowa State Census \citep{feigenbaum2018multiple}, 1918-19 BLS survey data \citep{feigenbaum2015intergenerational}, or New York City income tax data digitized from newspaper lists \citep{marcin2014essays}. The first two sources are ill-suited to the study of intergenerational mobility across different racial/ethnic groups, and the latter source is not publicly available.} 

It could also be the case that researchers are interested in measuring historical earnings differences along dimensions not featured in our lasso regressions. Since the variation in LIDO scores is entirely driven by the set of variables used to construct them, the specific set of possible predictors used is key. We purposefully excluded some possible adjustment variables to ensure that our LIDO scores could be used in every Census dating back to 1850. We also excluded certain variables we felt would lead to less accurate results. If there are substantial changes in the labor market outcomes of a specific demographic over time, perhaps because of composition changes, then adjustments specific to this demographic may yield unsatisfactory results.\footnote{See footnote \ref{immigrationnote} in reference to the exclusion of foreign-born status as a predictor.} Depending on the specific research question or data sources used, it may be the case that adjusting scores based on additional or alternative variables would lead to better estimates. We strongly encourage researchers to consider how our LIDO scores may or may not fit their needs, and we believe that our proposed cross-validated lasso approach could be fruitfully applied to help construct adjusted income scores using any number of alternative sets of predictors.

Studies that are primarily interested in occupational status, rather than earnings, may wish to retain the standard OCCSCORE. The 1950 OCCSCORE provides a ranking of occupations by earnings, and we find substantial persistence in these occupational rankings. Even for such studies, labor economists using modern data almost invariably use earnings instead of using OCCSCORE, and using LIDO scores would make the historical literature more comparable to the modern labor economics literature. For these reasons, we recommend the LIDO score as a complement rather than a substitute to the occupational income score.

\begin{singlespacing}
\bibliographystyle{econ}
\bibliography{biblio}
\end{singlespacing}

\begin{table}[ht]
\captionsetup{justification=centering}
\begin{center}
\caption{Earnings in the 1915 Iowa State Census}
\footnotesize
\begin{tabular}{lccc} 
\toprule
\multirow{2}{*}{Full sample} & Log of earnings & Log of 1950 OCCSCORE & LIDO score \\
%&  &  &   \\ %1950 OCCSCORE
 \cmidrule{2-4}
& (1) & (2) & (3) \\
\midrule
\multirow{2}{0.9in}{Black} & -0.389*** & -0.243*** & -0.422*** \\
 & (0.0378) & (0.0255) & (0.0232) \\
\multirow{2}{0.9in}{Female} & -0.516*** & 0.020* & -0.507*** \\
 & (0.0201) & (0.0106) & (0.0096) \\
\multirow{2}{0.9in}{Age} & 0.216*** & 0.033*** & 0.078*** \\
 & (0.0072) & (0.0045) & (0.0036) \\
\multirow{2}{0.9in}{Age$^2$} & -0.144*** & -0.008** & -0.069*** \\
 & (0.0070) & (0.0037) & (0.0030) \\
\multirow{2}{0.9in}{Urban} & 0.171*** & 0.332*** & 0.191*** \\
 & (0.0106) & (0.0060) & (0.0049) \\
\multirow{2}{0.9in}{Foreign born} & -0.146*** & -0.092*** & -0.088*** \\
 & (0.0186) & (0.0099) & (0.0077) \\
\midrule
Observations & 15,201 & 15,201 & 15,201 \\
$R^2$ & 0.152 & 0.125 & 0.316 \\
\bottomrule
\end{tabular}
\captionsetup{justification=justified}
\caption*{\footnotesize {\bf Notes:} Linear regressions of earnings measures on indicators for black, female, urban, and foreign-born status as well as a quadratic polynomial in age. Sample excludes those whose race is recorded as Missing, Mixed, or Asian (24 observations), those who are below the age of 20 or above the age of 70, those with missing occupation data, and those with zero or missing earnings. *** p$<$0.01, ** p$<$0.05, * p$<$0.1} 
\label{iowa_results}
\end{center}
\end{table}

\begin{table}
\caption{Estimates of Intergenerational Mobility}
\label{table:inter}
\begin{center}
\scalebox{.8}{
{
\def\sym#1{\ifmmode^{#1}\else\(^{#1}\)\fi}
\begin{tabular}{l*{6}{c}}
\hline
\multicolumn{7}{c}{Panel A: Mobility among whites} \\
\multicolumn{7}{c}{Dependent variable: log of son's OCCSCORE} \\
  & (1) & (2) & (3) & (4) & (5) & (6) \\
  &1850-1880 & 1860-1880 & 1880-1900 & 1880-1910 & 1880-1920 & 1880-1930 \\
\hline
Log of father's OCCSCORE&       0.402\sym{***}&       0.460\sym{***}&       0.544\sym{***}&       0.428\sym{***}&       0.403\sym{***}&       0.379\sym{***}\\
            &    (0.0219)         &    (0.0172)         &    (0.0127)         &    (0.0134)         &    (0.0150)         &    (0.0146)         \\
\(N\)       &        2804         &        3945         &        8354         &        7345         &        5687         &        5524         \\
\hline \\
[1em]
\multicolumn{7}{c}{Dependent variable: log of son's LIDO score} \\
  & (1) & (2) & (3) & (4) & (5) & (6) \\
  &1850-1880 & 1860-1880 & 1880-1900 & 1880-1910 & 1880-1920 & 1880-1930 \\
\hline
Log of father's LIDO score&       0.457\sym{***}&       0.439\sym{***}&       0.408\sym{***}&       0.367\sym{***}&       0.394\sym{***}&       0.387\sym{***}\\
            &    (0.0193)         &    (0.0154)         &    (0.0107)         &    (0.0105)         &    (0.0132)         &    (0.0138)         \\
\(N\)       &        2804         &        3945         &        8354         &        7345         &        5687         &        5524         \\
\hline \\
[1em]
\multicolumn{7}{c}{Panel B: Mobility among blacks} \\
\multicolumn{7}{c}{Dependent variable: log of son's OCCSCORE} \\
  & (1) & (2) & (3) & (4) & (5) & (6) \\
  &1850-1880 & 1860-1880 & 1880-1900 & 1880-1910 & 1880-1920 & 1880-1930 \\
\hline
Log of father's OCCSCORE& &&       0.174\sym{***}&       0.161\sym{**} &      0.0825         &       0.184\sym{*}  \\
            &&&    (0.0514)         &    (0.0581)         &    (0.0698)         &    (0.0786)         \\
\(N\)       &&&         520         &         375         &         250         &         204         \\
						
\hline \\
[1em]
\multicolumn{7}{c}{Dependent variable: log of son's LIDO score} \\
  & (1) & (2) & (3) & (4) & (5) & (6) \\
  &1850-1880 & 1860-1880 & 1880-1900 & 1880-1910 & 1880-1920 & 1880-1930 \\
\hline
Log of father's LIDO score&&&       0.600\sym{***}&       0.541\sym{***}&       0.562\sym{***}&       0.506\sym{***}\\
            &&&    (0.0436)         &    (0.0600)         &    (0.0954)         &    (0.0969)         \\
\(N\)       &&&         520         &         375         &         250         &         204         \\
\hline

\end{tabular}
}}
\end{center}
\footnotesize{  { \bf Notes:} Data are from the IPUMS linked data files. These data are from 1\% samples of the 1850, 1860, 1900, 1910, 1920 and 1930 Censuses linked to the 1880 complete count. The sample is restricted to those who during the first Census year were children of the household head, male, and no older than 15 years old. Standard errors are in parentheses.
* $p<0.1$; ** $p<0.05$; *** $p<0.01$. }
\end{table}

\end{document}